\documentclass{pasa}%
\usepackage{natbib}
\usepackage{soul,color}
\def\url#1{\expandafter\string\csname #1\endcsname}

\def\actaa{Acta. Astronom.} %
\def\aj{AJ}%
\def\apj{ApJ}%
\def\apjl{ApJ}%
\def\apjs{ApJS}%
\def\aap{A\&A}%
\def\mnras{MNRAS}%
\def\pasa{PASA}%
\def\pasp{PASP}%
\def\nat{Nature}%
\title[Bulge Dynamics Review]{Was the Milky Way Bulge Formed From The Buckling Disk Instability,  Hierarchical Collapse, Accretion of Clumps, or All of the Above?}
\author[Nataf]{David M. Nataf$^1$\thanks{dnataf1@jhu.edu, david.nataf@gmail.com}\\
\affil{$^1$ Center for Astrophysical Sciences and Department of Physics and Astronomy, The Johns Hopkins University, Baltimore, MD 21218.}%
}
\jid{PASA}
\doi{10.1017/pas.\the\year.2015}
\jyear{\the\year}

\begin{document}%
\begin{abstract}
The assembly of the Milky Way bulge is an old topic in astronomy, one now in a period of renewed and rapid development. That is due to tremendous advances in observations of bulge stars; motivating observations of both local and high-redshift galaxies; and increasingly sophisticated simulations. The dominant scenario for bulge formation is that of the Milky Way as a nearly pure disk galaxy, with the inner disk  having formed a bar and buckled. This can potentially explain virtually all bulge stars with [Fe/H] $\gtrsim -1.0$, which comprise 95\% of the stellar population. The evidence is the incredible success in N-body models of this type in making non-trivial, non-generic predictions, such as the rotation curve and velocity dispersion measured from radial velocities, and the spatial morphologies of the peanut/X-shape and the long bar. The classical bulge scenario, whereby the bulge formed from early dissipative collapse and mergers, remains viable for stars with [Fe/H] $\lesssim -1.0$ and potentially a minority of the other stars. A classical bulge is expected from $\Lambda$-CDM cosmological simulations, can accentuate the properties of an existing bar in a hybrid system, and is most consistent with the bulge abundance trends such as [Mg/Fe], which are elevated relative to both the thin and thick disks. Finally, the clumpy-galaxy scenario is considered, as it is the correct description of most Milky Way precursors given observations of high-redshift galaxies. Simulations predict that these star-forming clumps will sometimes migrate to the centres of galaxies where they may form a bulge, and galaxies often include a bulge clump as well. They will possibly form a bar with properties consistent with those of the Milky Way, such as the exponential profile and metallicity gradient. Given the relative successes of these scenarios, the Milky Way bulge is plausibly of composite origin,  with a classical bulge and/or inner halo numerically dominant for stars with [Fe/H] $\lesssim -1.0$, a buckling thick disk for stars with $-1.0 \lesssim \rm{[Fe/H]]} \lesssim -0.50$ perhaps descended from the clumpy galaxy phase, and a buckling thin disk for stars with  [Fe/H]$\gtrsim -0.50$. Overlaps from these scenarios are uncertain throughout. 
\end{abstract}
\maketitle%
\section{INTRODUCTION}
\label{sec:intro}

Formation scenarions for the Galactic bulge (and bulges in general) have been around for some time. 

\citet*{1962ApJ...136..748E} suggested the scenario of early, rapid, monolithic formation by dissipative collapse, this is now referred to as the ``classical bulge" scenario. The hierarchical merging of smaller objects, typical of the earliest phases of galaxy formation as predicted by $\lambda$-CDM cosmological simulations \citep{2010ApJ...708.1398T,2011ApJ...729...16K}, is now often included as part of the definition of ``classical bulges" \citep{2005MNRAS.358.1477A}. It is also known that bars can form from the dynamical evolution of a disk galaxy via the ``buckling instability" scenario, where the bulge would be predominantly composed of disk stars now on bar orbits \citep{1978ApJ...223..811M,1979ApJ...227..785M,1979AJ.....84..585H}, for example the x1 orbital family \citep{2002MNRAS.333..847S}. Finally, the migration of star-forming clumps toward the centres of disk galaxies due to dynamical friction emerges naturally from simulations, this is the ``clump-origin" scenario \citep{1998Natur.392..253N,1999ApJ...514...77N}.   

Given the wide availability of plausible models, it is no surprise that it has long been acknowledged that the formation of the bulge may be a composite process. For example, in a previous review (of bulges in general), \citet{2000bgfp.conf..413C} wrote:
\begin{quotation}
 \vspace{-0.40cm}   \noindent \textit{A fraction of bulges could have formed early (at first collapse); then secular dynamical evolution enrich them; in parallel, according to environment, accretion and minor mergers contribute to raise their mass.} 
    \end{quotation}
   \vspace{-0.40cm}   The first and third process mentioned are the two components of the classical bulge scenario, the second is the buckling instability scenario. Relative contributions of the different scenarios of bulge formation were not assigned precise bounds, given that this is an intrinsically difficult problem. 
   
   Astronomers need an accurate, precise and thorough census of the current Milky Way bulge before we can seriously consider disentangling its history.    
   
   That census is now slowly, but surely, becoming available. These include ground-based photometric surveys such as OGLE-IV in the optical \citep{2015AcA....65....1U}, $VVV$ in the near-infrared \citep{2012A&A...537A.107S} and  \textit{Wide-field Infrared Survey Explorer (WISE)} in the mid-infrared \citep{2010AJ....140.1868W};  Deep, multi-wavelength \textit{Hubble Space Telescope (HST)} photometry which measures the main-sequence and thus constrains ages \citep{2010ApJ...725L..19B}. At the spectroscopic end, surveys such as BRAVA \citep{2012AJ....143...57K}, ARGOS \citep{2013MNRAS.428.3660F},  Gaia-ESO \citep{2014A&A...569A.103R}, EMBLA \citep{2014MNRAS.445.4241H}, APOGEE \citep{2016ApJ...819....2N}, BRAVA-RR \citep{2016ApJ...821L..25K}, and GIBS \citep{2017A&A...599A..12Z}, collectively provide kinematics and sometimes detailed chemistries for tens of thousands of bulge stars. 
   
   Concurrent with this, there have been vast improvements in the theory and simulations as well. For example, the N-body simulation of \citet{2017MNRAS.467L..46A} contains 17.5 million particles, whereas that of \citet{1979ApJ...227..785M} had 96,017 particles. The level of detail, and sophistication of current models allows more robust analysis. Further, more questions are being asked of models, such as how  a classical bulge will evolve when embedded within a massive disk \citep{2012MNRAS.421..333S},  where younger stars might be distributed within a bar \citep{2016arXiv161109023D}, or how the separation between the two arms of an X-shape will appear as a function of direction \citep{2015MNRAS.447.1535N}. These questions, largely motivated by the observations, should yield more predictive and discriminatory power in evaluating models, with the caveat that it is scientifically misguided to expect a perfect quantitative match between a simulation and an actual galaxy. 
   
   In this review,  three scenarios for bulge formation are discussed as well as the current evidence in their favour. 
   
      \section{THE MILKY WAY AS A NEARLY PURE DISK GALAXY}
This is the scenario where the Milky Way bulge is largely or nearly entirely a bar. This bar would have first formed from a disk, and then extended vertically by one or more buckling instability episodes. This scenario arguably has the most evidence in its favour.

The evidence presented is that:
\begin{itemize}
\item Pure disk galaxies are observed to exist in the local universe, and the bar buckling process occurs naturally in N-body simulations.
\item The radial velocity measurements from large spectroscopic surveys are consistent with the theoretical predictions from buckling disk models.
\item The peanut/X-shaped morphology for the distribution of bulge stars is a prediction of these models, and is constrained to represent a large fraction of the bulge mass. 
\item The morphology of the long bar is another specific, precise prediction of these models.
\item An old argument against this theory, that of the metallicity gradient, has been shown to be invalid as it can also be produced by the buckling instability model.
\end{itemize}

\subsection{Pure disk galaxies exist, and simulations show that they can naturally evolve to be barred galaxies}
   Pure disk galaxies exist. \citet{2010ApJ...723...54K} obtained \textit{HST} photometry of six nearby galaxies and measured their surface brightness profiles. They found an upperbound on the total (classical bulge + pseudobulge) stellar mass fraction of $\sim$3\% of the total disk mass. They also took an inventory of galaxies within 8 Mpc with $v_{circ} > 150\,\, \rm{km\, s}^{-1}$, and found that 11 of 19 showed no evidence for a classical bulge, and four of the remaining eight may contain classical bulges contributing 5\% - 12\% of the stellar mass. It is the case, however surprisingly, that pure disk galaxies are common in the low redshift universe, and thus it is plausible for the Milky Way to be one as well.  
   
   Simulations of disk galaxies consistently show that bars and subsequently buckling bars can  occur spontaneously in disk galaxies (e.g. \citealt{1981A&A....96..164C,1991Natur.352..411R,2005MNRAS.358.1477A,2006ApJ...637..214M,2014MNRAS.437.1284Q}). The rate of evolution of the bar is sensitive to various variables, such as the ratio of the disk to halo mass \citep{1981A&A....96..164C}. The bar may have a north-south asymmetry at first, but this rapidly dissipates \citep{1991Natur.352..411R}. Overtime, the bars tend to grow in vertical extent, for example via multiple, recurrent buckling episodes (\citealt{2006ApJ...637..214M}, see Figure 3). 
   
The combination of these two facts, that pure disk galaxies are common in the local universe, and that simulations predict that pure disk galaxies can spontaneously evolve to have bars, render it a plausible model for the Milky Way's bulge as well. The necessary initial conditions -- a disk galaxy -- are ubiquitous, and the necessary evolution is natural. 

\subsection{Radial velocity measurements from large spectroscopic surveys}
The Milky Way is a barred galaxy  \citep{1995ApJ...445..716D}, however it has not been clear until recently how quantitatively dominant the bar is relative to the hypothetical classical bulge contribution. The advent of large spectroscopic surveys has allowed us to disentangle the bulge into plausible subcomponents. 

\citet{2010ApJ...720L..72S} fit a suite of N-body models to bulge radial velocity data from the BRAVA survey, specifically the mean radial velocity (and thus rotation curve) and radial velocity dispersion as a function of longitude and latitude. They recover a best-fit viewing angle, betweeen the bar's major axis and the line of sight between the Sun and the Galactic centre, of $\alpha_{\rm{Bar}}=20^{\circ}$, consistent with more recent measurements. More significantly, they  show that models with significant classical bulges, with parameters defined to lie on the fundamental plane of ellipticals and bulges \citep{2009ApJS..182..216K}, are not well-fit by the data. They constrain the classical bulge mass of the Milky Way to be no more than 8\% of the total mass of the disk. 

\citet{2013MNRAS.432.2092N} analyzed the data from the larger ARGOS survey, specifically 17,400 red giants with best-fit parameters including  $R_{GC} \leq 3.5$ Kpc, which were spread over 30 degrees of longitude and 20 degrees of latitude. They find several features in the data consistent with N-body expectations for disk origin to the bulge. Bulge stars located between 5 and 10 degrees from the plane ($0.7\, \rm{Kpc} \lesssim |z| \lesssim 1.5\, \rm{Kpc}$) are cylindrically rotating, with the exception of the 5\% of stars with [Fe/H] $\leq -1.0$. They suggest that those stars are the inner extensions of the halo and the metal-weak thick disk. It is also the case that the thin disk includes very, very few stars with [Fe/H] $\leq -0.50$ \citep{2017arXiv170504349D}, so it was always an implausible origin source for the metal-poor component of the bulge. 

\citet{2015A&A...577A...1D} compared the ARGOS data to three different N-body models and found that the bulge could not be explained as having resulted from the evolution of a pure thin disk galaxy, and thus a joint origin with the thick disk is likely required as well. This may appear as requiring too many free parameters from models, but this is a parameter that exists in nature, the Milky Way is not a pure thin disk galaxy, it also has a thick disk. \citet{2015A&A...577A...1D} show that if the bulge was purely due to a thin disk buckling event, the curve of mean radial velocity versus longitude would decrease in amplitude, and the velocity dispersions would decrease, with increasing metallicity. The decrease in the amplitude of the rotation curve is seen for stars satisfying [Fe/H] $\gtrsim -0.50$, but not for lower metallicities. The velocity dispersions do decrease, but not as quickly as predicted from a pure thin disk model. \citet{2015A&A...577A...1D} suggest that the solution to this issue is that bulge stars with [Fe/H] $\lesssim -0.50$ formed from the thick disk, rather than from the more metal-poor component of the thin disk. 

\citet{2016ApJ...832..132Z} investigated the skewness (third moment) and kurtosis (fourth moments) of the radial velocity distribution functions in APOGEE data toward the bulge. The correlation between skewness and mean velocity (first moment), a known diagnostic of bars, is only observed for the more metal-rich fraction of the stars, defined in that paper as stars with [Fe/H] $\gtrsim -0.40$. The data have a flat kurtotsis, Kurt(V) $\approx 0$, consistent with that expected from models. The data have slightly higher velocity dispersion and slightly lower skewness than expected from  N-body models of a simple single disk galaxy undergoing buckling. 

The recent review of \citet{2016PASA...33...27D} goes over these lines of evidence in greater detail than possible here, and concludes that bulge stars with [Fe/H] $\gtrsim -1.0$ formed from buckling thick and thin disks.

\subsection{The peanut/X-shaped spatial distribution of bulge stars}

\begin{figure}
\begin{center}
\includegraphics[totalheight=0.19\textheight]{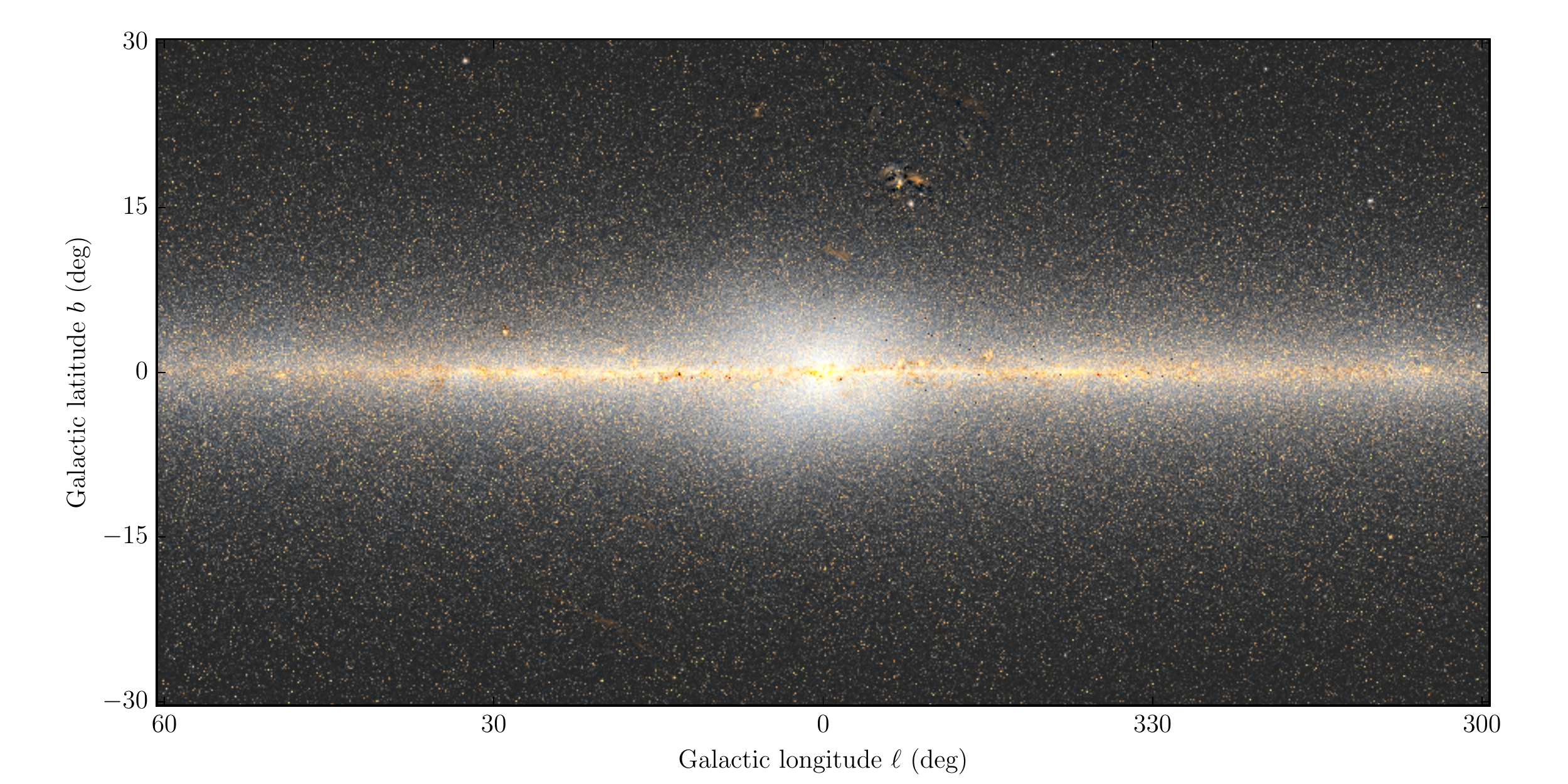}
\caption{Figure 1 from \citet{2016AJ....152...14N}, the X-shaped bulge is unambiguous in the integrated mid-infrared photometry of the Milky Way.}
\label{NessLang}
\end{center}
\end{figure}

\begin{figure}
\begin{center}
\includegraphics[totalheight=0.40\textheight]{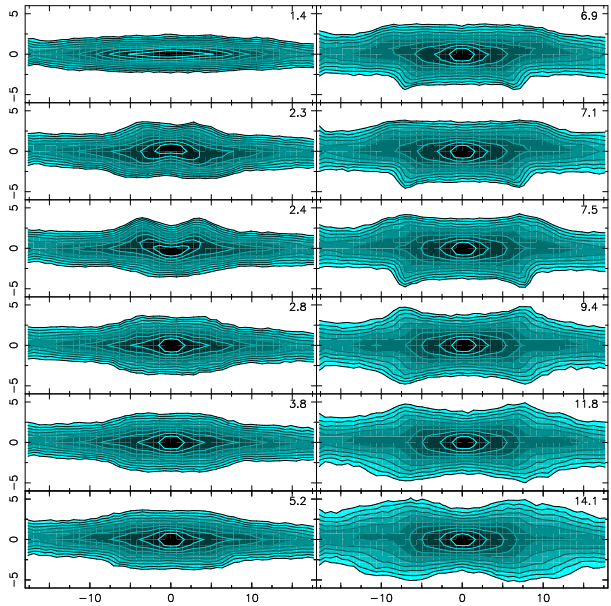}
\caption{Figure 3 from \citet{2006ApJ...637..214M}. An X-shaped bulge is a natural outcome of N-body models of disk galaxies undergoing the buckling instability.}
\label{MV2006}
\end{center}
\end{figure}

The sightline-dependent bifurcation of the apparent magnitude distribution of red clump stars (comprising $\sim$99\% of bulge horizontal branch stars, see \citealt{2013ApJ...769...88N}) was first reported by \citet{2007MNRAS.378.1064R}. They speculated that it might be due to a distinct population lying at a different distance. \citet{2010ApJ...721L..28N}  and \citet{2010ApJ...724.1491M} showed that the double peak was spread across a large swath of the bulge, at large separations ($|b| \gtrsim 5^{\circ}$) from the plane.  \citet{2010ApJ...724.1491M} argued convincingly, that the feature had to be due to an X-shaped bulge (a very strong peanut shape), a feature of strong bars in N-body models.  

Subsequent analyses confirmed the claim. \citet{2012ApJ...756...22N} found that the bright and faint red clumps had the same difference in mean radial velocity as expected from N-body models.  \citet{2012ApJ...757L...7L} showed that the N-body model of \citet{2010ApJ...720L..72S}, already demonstrated to be a good match to radial velocity observations, also includes an X-shape.  The relative brightness, and the dependence of the relative number counts on direction, was qualitatively similar to that seen in the data. \ \citet{2013A&A...555A..91V} showed that the proper motion distributions were consistent. \citet{2015MNRAS.447.1535N} showed that precision measurements of the brightness difference, relative number counts, and total number density of stars in the data from OGLE-III \citep{2008AcA....58...69U} was qualitatively matched by N-body models from \citet{2003MNRAS.341.1179A} and \citet{2010ApJ...720L..72S}. An unambiguous image of the X-shaped bulge in integrated mid-infrared photometry of the Milky Way is shown in Figure \ref{NessLang}, and the development of one from an N-body model is shown in Figure \ref{MV2006}.

Precision modelling of the photometric data by \citet{2013MNRAS.435.1874W} has been followed by detailed dynamical modelling by \citet{2015MNRAS.450L..66P,2015MNRAS.448..713P,2017MNRAS.465.1621P}. A suite of N-body models coarsely consistent with Milky Way constraints were adjusted to be more consistent using the made-to-measure method \citep{1996MNRAS.282..223S}, whereby weights of the particles in the model are shifted up or down to match the observational data. The end models always contain a significant X-shape, comprising 40-50\% of the stellar mass of the bulge, but only dominant for stars with [Fe/H] $\gtrsim -0.50$. That is not a repetition of the results mentioned in the previous subsection, as \citet{2015MNRAS.450L..66P,2015MNRAS.448..713P,2017MNRAS.465.1621P} also constrained their N-body models to match the spectroscopic data from the APOGEE survey, which goes much closer to the plane \citep{2016ApJ...819....2N}, and the global photometric parameters of the bulge which trace the number density and distance distribution function of stars   \citep{2013MNRAS.435.1874W}. 

Reiterating, this is an impressive series of observational tests for models to have passed. First, the alternative hypotheses (RGB bump, metallicity differences, etc) do not work. Second, N-body model predictions qualitatively match the radial velocity offsets, the proper motion measurements, the mean brightness difference, and the dependence of the relative and total number counts as a function of direction. This picture survives, and is in fact dominant for stars with [Fe/H] $\gtrsim -0.50$, when models are required to match observations from several different surveys. 

\subsection{ The morphology of the long bar is another specific, precise prediction of these models.}
An issue of Galactic modelling in the last two decades is that of the long bar of the Milky Way \citep{2005ApJ...630L.149B,2007A&A...465..825C,2008A&A...491..781C}. Evidence has accumulated from multiple investigations of a long, in-plane bar, with a half-length of $\sim$4 Kpc, and an orientation angle of $\alpha_{\rm{Bar}}=45^{\circ}$. This represented a challenge to Galactic structure studies, as the orientation angle is far larger than the value found for the triaxial bulge toward the inner few Kpc, and the morphology is not consistent with the traxial ellipsoid models used to fit for the bulge \citep{1995ApJ...445..716D,1997ApJ...477..163S}. It appears as a double-barred system where the two bars are of comparable length and are not aligned, which is an unstable configuration \citep{2012EPJWC..1906004A}.

A suggestion to this resolution was independently proposed by \citet{2011ApJ...734L..20M} and \citet{2012EPJWC..1906004A}. Their idea was the analytical triaxial ellipsoid models then widely used to model the bulge were only (somewhat) suitable for the peanut/X-shaped component, whereas bars in simulations often have long, and thin extensions. This hypothesis made a specific prediction, that the difference in measured angle ${\Delta}\alpha_{\rm{Bar}}$, then around 20 degrees, was due to observational errors. A smaller difference might remain if the ends of the bar develop interactions with spiral arms  \citet{2011ApJ...734L..20M}.

This prediction was confirmed by \citet{2015MNRAS.450.4050W}, who combined photometry from Spitzer-GLIMPSE \citep{2005ApJ...630L.149B}, 2MASS \citep{2006AJ....131.1163S}, VVV \citep{2012MNRAS.422.1902I}, and UKIDSS \citep{2008MNRAS.391..136L} to make a global map of the bulge.  The long bar had an angle of $\alpha_{\rm{Bar}}=28-33^{\circ}$, consistent with the values of $\alpha_{\rm{Bar}}=27 \pm 2^{\circ}$  and  $\alpha_{\rm{Bar}} \approx 29^{\circ}$ then recently measured by \citet{2013MNRAS.435.1874W} and \citet{2013MNRAS.434..595C} respectively for the inner peanut/X-shaped component. 

Thus, the issue of the long bar, which was previously a challenge to Galactic structure models, ended up being a triumph. The long bar is not only not surprising, but expected from simulations of buckling disk galaxies. Models predicted that the difference in orientation angle should shrink with better analysis, and it did. 
   
   \subsection{ The metallicity gradient can also be reproduced by buckling disk}
The Galactic bulge has a vertical metallicity gradient, with stars further from the plane having a lower mean metallicity. \citet{1995AJ....110.2788T} used near-IR photometry and estimated a gradient in ${\nabla}$[Fe/H] $= -0.06\pm0.03$ dex/deg or $-0.43\pm0.21$ dex/kpc between $b=-3^{\circ}$ and $b=-12^{\circ}$. The metallicity gradient was independently and concurrently confirmed by \citet{1995MNRAS.277.1293M}, and since confirmed numerous times with spectroscopic data (e.g. \citealt{2008A&A...486..177Z,2011ApJ...732..108J}) 

This was argued by \citet{1995MNRAS.277.1293M} to be evidence for a dissipative collapse, as models of spheroid formation predicted a radial metallicity gradient \citep{1984ApJ...286..403C}. However, \citet{1995MNRAS.277.1293M} acknowledged that  ``\textit{the alternative interpretation that the gradient itself is caused by the mixing of different components in the inner Galaxy cannot be ruled out.}''

That second point, that the bulge has different components with both different metallicity distribution functions and scale heights, is very much the case, as should be clear from the literature evidence summarized in this review. Further, it also turns out that a metallicity gradient for the bulge can emerge from a pure disk galaxy. \citet{2013ApJ...766L...3M} evolved an N-body model of a pure disk galaxy which included an initial radial metallicity gradient. They chose $\rm{[M/H]}(R)=+0.60-0.40(R)$/Kpc, where ``R'' is the galactocentriuc radius of particles at the start of the simulation. Given that particles from initial radii are scattered to different orbits, a metallicity gradient emerges which is qualitatively consistent with that observed for the whole bulge -- there is impressive agreement with the global photometric metallicity maps from \citet{2013A&A...552A.110G}, which are derived from the morphology of the red giant branch. That said, it is worth noting that the metallicity gradient required by \citet{2013ApJ...766L...3M} is extremely large. 

This result is not altogether surprising as it has been known for a while that dynamical mixing processes do not erase gradients. For example, \citet{1980MNRAS.191P...1W} demonstrated that simulations of mergers were expected to reduce, but not erase, gradients in metallicity. \citet{2014A&A...563A..49S} finds that the metallicity gradients in interacting galaxies, measured in terms of oxygen abundance of HII regions versus effective radius, is reduced by $\sim 1/2$ relative to non-interacting galaxies. 

The metallicity gradient of the bulge, first largely argued to be evidence for dissipative collapse and thus evidence against a disk-galaxy origin, turns out not to be discriminating. A metallicity gradient can be matched by models of dissipative collapse, pure disk galaxies undergoing the buckling instability, a combination of the two, or as we will see, the clump-origin bulge scenario \citep{2012MNRAS.422.1902I}. The existence of a metallicity gradient, by itself, yields no significant constraints bulge formation. 

\subsection{Caveat against the bar scenario: Gas fractions and the challenge of initial conditions}

The properties and evolution of bars in N-body simulations of pure exponential disks are a well-researched subject. However, real galaxies (such as the Milky Way) contain gas, and generally contained more gas in the past. 

\citet{2013MNRAS.429.1949A} investigated the relative predicted properties of bars in N-body simulations with and without gas. They found that the gas-rich galaxies remain axisymmetric for longer. When they do develop bars, they do so at a slower rate, and end up much weaker. Their Figure 7 show that bar strengths in gas-rich galaxies are predicted to end up lower even after the gas has been depleted.  In contrast, as discussed in this section, the Milky Way has a very strong bar. 

Indeed, the bar fraction is considerably lower at high redshift. \citet{2014MNRAS.438.2882M} finds that it drops to 11\% at $z=1$, corresponding to a lookback time of 7.8 Gyr. \citet{2014MNRAS.445.3466S} also estimates a bar-fraction of 11\%, in the redshift range $0.50 < z < 2.0$.  By that time most Milky Way bulge stars were already formed \citep{2011ApJ...735...37C,2017arXiv170202971B} and thus they already have a kinematic distribution.

The combination of these issues, should give pause to the notion that a bar+buckling instability from a pure exponential disk is independently sufficient to explain the inner Milky Way's dynamics. Separately from the issue of gas weakening bars, subsequent sections of this review will discuss how a minor classical bulge can actually strengthen the bar, and why the Milky Way is expected to be a former ``clumpy galaxy" which has implications for bulge formation.

   \section{THE CLASSICAL BULGE SCENARIO}
   The classical bulge scenario is one where the bulge is formed a combination of early, dissipative collapse and accretion of objects via minor or major mergers. 
      
   The evidence we present is that:
\begin{itemize}
\item It is expected from theory.
\item The metal-poor bulge stars are kinematically consistent with a classical bulge behaviour. 
\item  The theoretical interaction between classical bulges and buckling disks is consistent with observations.
\item The bulge trends in the $\alpha$-elements are not consistent with those of the disk. 
\end{itemize}

\subsection{A classical bulge is expected from theory}

The most obvious advantages of this scenario are that it is predicted by straightforward theory of gravitational, dissipative collapse \citep{1962ApJ...136..748E}, and hierarchical clustering in a cold dark matter universe \citep{1978MNRAS.183..341W,1993MNRAS.264..201K}. 

 \citet{1978MNRAS.183..341W} proposed that most of the matter in the universe condensed early into small ``dark'' objects. They suggested a model with $\Omega_{m}=0.20$ and dark matter making up 80\% of the matter, impressively similar to the modern values \citep{2014A&A...571A..16P}.  Within this picture, the ``pure disk galaxy" that has particles and no gas is not a viable initial condition, as the universe and star formation begin with larger number of small haloes that coalesce and accrete additional small haloes. 
 
 As far as current, more up-to-date models are concerned, a classical bulge can be considered a requirement. \citet{2011ApJ...729...16K} simulate 150 galaxies from various, cosmologically-motivated initial conditions. They formed disk structures in 48 of their galaxies, including 5 galaxies that had masses comparable to the Milky Way. Though some of their simulated galaxies had small classical bulges, none completely lacked a classical bulge. It is a concern that many local galaxies are found not to have classical bulges locally \citep{2010ApJ...723...54K}, but one can argue against that by saying that information is lost when looking at those galaxies in integrated light. That cannot be argued for the Milky Way. 
 
 \subsection{The kinematics of metal-poor bulge stars }
 
 \begin{figure}
\begin{center}
\includegraphics[totalheight=0.32\textheight]{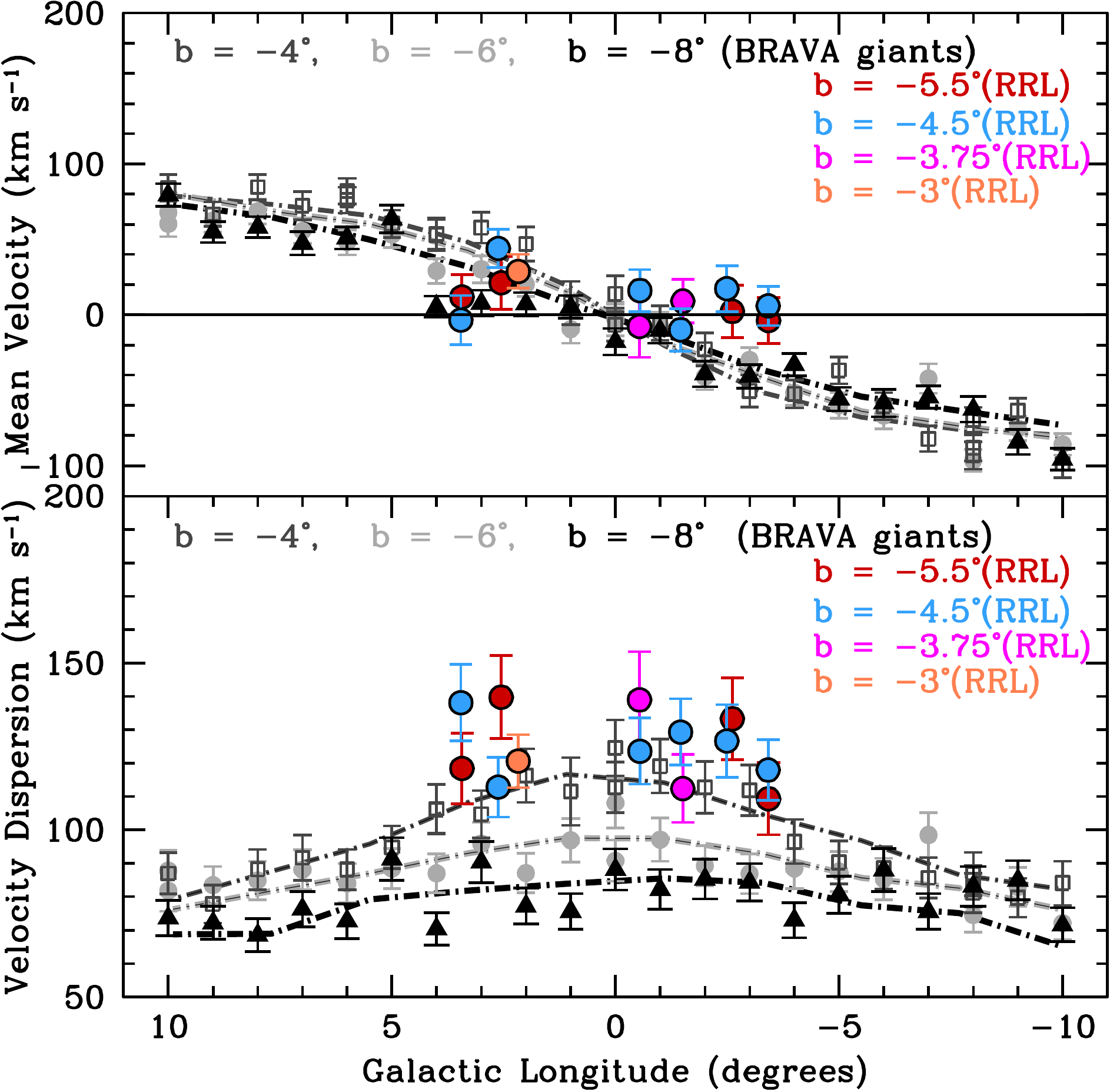}
\caption{Left panel of Figure 2 from \citet{2016ApJ...821L..25K}. RR Lyrae stars show null or negligible Galactic rotation, as well as very high velocity dispersion, in contrast to the majority of bulge stars.}
\label{Kunder2016}
\end{center}
\end{figure}

There has been much written in this review of the spectacular consistency between N-body models of buckling disks and the dynamics of bulge stars with [Fe/H] $\gtrsim -1.0$ or $\gtrsim -0.50$, depending on the study. These consistencies are not found for the metal-poor bulge stars. For example, the Galactic bar is either null \citep{2013ApJ...776L..19D} or weak \citep{2012ApJ...750..169P,2015ApJ...811..113P} in the RR Lyrae stars, variable stars of standardizable distance that can be used to directly probe the metal-poor bulge. \citet{2013ApJ...776L..19D} investigated 7,663 fundamental-mode RR Lyrae in $I$ and $K_{s}$ bands, and found that they do not trace a strong bar, but rather a more spheroidal and centrally concentrated distribution. No correlation is observed between dereddened distance and longitude, a correlation detected at high significance in the more metal-rich red clump stars which trace a bar \citep{1997ApJ...477..163S}. 
    
    \citet{2012ApJ...750..169P} and \citet{2015ApJ...811..113P} obtained different results in their analysis of 16,836 and 27,258 RR Lyrae in the OGLE-III and OGLE-IV surveys respectively. They do find a bar, but it is kinematically hotter than the bar measured in red clump stars, and does not show a peanut/X-shape at large separations from the plane. 
    
   Analysis of RR Lyrae kinematics further these findings. \citet{2016ApJ...821L..25K} analyze spectroscopic data for 947 RR Lyrae as part of the ongoing BRAVA-RR survey. These RR Lyrae, measured toward  $|l| \lesssim 4^{\circ}$ and $-6^{\circ} \lesssim b \lesssim -3^{\circ}$ show higher velocity dispersions and weaker rotation than the metal-rich M-giants studied as part of the BRAVA survey. The velocity dispersion is $\sim$15\% higher, and the rotation is null or negligible. When the RR Lyrae are split into two metallicity bins with [Fe/H]$=-$0.75 marking the bifurcation point, no difference in rotation is measured though the metal-poor RR Lyrae do have a higher velocity dispersion. The RR Lyrae radial velocity measurements are shown in Figure \ref{Kunder2016}.
    
    A further issue with the RR Lyrae is the very fact that they are RR Lyrae. The bulge RR Lyrae are measured to have a mean metallicity of [Fe/H]$\approx -1.0$, whereas metallicities of [Fe/H]$\approx -1.60$ are more typical of RR Lyrae in the globular clusters and in the halo \citep{2010ApJ...708..698D}. More metal-rich stars have a higher turnoff mass at fixed age, and thus for these horizontal branch stars to be on the instability strip at a higher metallicity, they need to have a lower turnoff mass by other means, and thus likely a greater age. \citet{1992AJ....104.1780L} estimated that the bulge RR Lyrae had to be ${\Delta}t \sim 1.3 \pm 0.30$ Gyr older than halo stars, for stars formed with metallicities $-1.5 \lesssim \rm{[Fe/H]} \lesssim -1.0$. A bulge which is older than the halo is more consistent with a classical bulge then say, a buckling disk origin. 
    
This argument also applies to bulge globular clusters. For example, NGC 6522, has a blue horizontal branch with a mean metallicity of [Fe/H]$=-1.0$ \citep{2009A&A...507..405B}, necessitating an extremely old age. This is a consistent pattern of bulge globular clusters \citep{2006A&A...449..349B,2007AJ....134.1613B}. This information, combined with that of the RR Lyrae, strongly suggests that bulge stars of metallicity [Fe/H]$=-1.0$ are the oldest or among the oldest stellar populations in the Galaxy, and thus were around prior to there being a massive disk that could form a bar. 
        
    \subsection{The theoretical interaction between classical bulges and buckling disks is consistent with observations.}
    There is relatively sparse research on the theoretically predicted interaction of a central concentration of stars on a buckling disk, but what research is available now turns out to be consistent with the data. This is true both of the predictions of the classical bulge behaviour and the bar behaviour. 
    
    \citet{2002MNRAS.330...35A} studied three different N-body models of disk galaxies with variable initial central concentrations. The disk mass and disk-to-halo mass ratios are fixed. Two models have a non-centrally concentrated halo, one of those two also has a bulge. A third model has a centrally concentrated halo, but no bulge. Within this suite of models, the galaxy with a non-centrally concentrated halo and no bulge ends up forming the weakest bar, it does not develop cylindrical rotation, and its boxy shape does not evolve to an X-shape. Interestingly, the model with a centrally concentrated halo but no bulge ended up with the strongest bar. This is suggestive that greater central concentration was needed given the exceptionally strong bar of the Milky Way, though in and of itself it is not sufficient as more models would be needed for a more convincing picture. 
    
    Conversely, \citet{2012MNRAS.421..333S,2013MNRAS.430.2039S,2016A&A...588A..42S}  have investigated how a joint classical bulge / buckling disk origin impacts the development of the bulge by means of N-body models. \citet{2012MNRAS.421..333S} finds that a small (7\% of the total dis mass) classical bulge can pick up angular momentum from the larger rotating bar, and thus even develop into a triaxial object with cylindrical rotation. \citet{2013MNRAS.430.2039S} show that the composite bulge always ends up rotating cylindrically, but may have deviations from cylindrical rotation at specific moments in its evolution. Curiously, they find that the final size of the composite bulges are reduced if the initial classical bulge has its own angular momentum. \citet{2016A&A...588A..42S}  extend their prior results to show that even massive classical bulges can pick up as much specific angular momentum as low-mass classical bulges, but that the resulting rotation is non-cylindrical. All composite systems eventually form a boxy/peanut bulge. 
    
       \citet{2017MNRAS.464L..80P} model the evolution of a disk galaxy with a halo via an N-body model and find that the properties of the RR Lyrae are consistent with that of an inner halo, specifically their number density, their slow rotation, the lack of a peanut/X-shape in their spatial distribution, and their higher velocity dispersion. Within their model, only 12\% of RR Lyrae end trapped on bar-like orbits, which can be considered a prediction that the fraction will be very small once proper motions are available and once astronomers can compute orbits for these stars. 
       
       It been established in previous sections of this review that observations and analysis thereof have ruled out a predominantly classical bulge for the Milky Way bulge. However, the N-body models are also consistent with either a centrally concentrated inner halo, or an additional classical bulge in addition to that. Such a feature could help explain the Milky Way's very strong bar as well as the behaviour of low-metallicity bulge stars. Further analysis, and more data of metal-poor bulge stars are needed.

       \subsection{The bulge trends in the $\alpha$-elements are not consistent with those of the thin and thick disks. }
              
       The abundances of the $\alpha$-elements relative to iron, [$\alpha$/Fe], are a historic argument for a distinct formation to the bulge and the disk. That is because the [$\alpha$/Fe] abundance ratios trace the efficiency of star formation and possibly the initial mass function of a stellar population \citep{1986A&A...154..279M}. A higher level of [$\alpha$/Fe] can be due to a lower contribution to the chemical enrichment of interstellar gas from type Ia SNe  (which take longer to form), and thus star formation would need to be faster. 
       

The landmark study of \citet{1994ApJS...91..749M} found that the trends for magnesium and titanium were enhanced by $\approx$0.30 dex relative to the solar neighbourhood trends over the full range of [Fe/H], whereas the trends for calcium and silicon were consistent with those of the disk. These different ratios suggested a dfferent origin, and \citet{1994ApJS...91..749M} said it may reflect a common enrichment process between bulges and ellipticals. \citet{2007ApJ...661.1152F} compiled a more sophisticated analysis of a higher resolution, higher signal-to-noise sample. They found that the bulge has an magnesium trend elevated by $\sim$0.30 dex relative to the disk, whereas the trends of oxygen, silicon, calcium, and titanium are slightly elevated relative to the disk. The bulge also has higher aluminum abundances at fixed iron abundance.  \citet{2007ApJ...661.1152F} conclude that the relative abundance offsets between the bulge and the disk are inconsistent with models where the bulge forms from the buckling of the disk. Separately,  \citet{2007ApJ...661.1152F} found that the metal-poor bulge stars also show higher mean abundances of silicon, calcium, and titanium than the halo. They concluded that the metal-poor bulge stars could not have formed from gas with the present-day halo composition. 

The discussion shifted with the work of \citet{2010A&A...513A..35A}. That investigation used high-resolution of optical spectra of 25 bulge giants and 55 comparison giants and analyzed abundances in a homogeneous manner to minimize systematic offsets. Their results were that metal-poor bulge stars ([Fe/H] $\lesssim -0.50$) have the same abundances as the thick disk, and more metal-rich stars have the same abundances as the thin disk.

\begin{figure}
\includegraphics[totalheight=0.17\textheight]{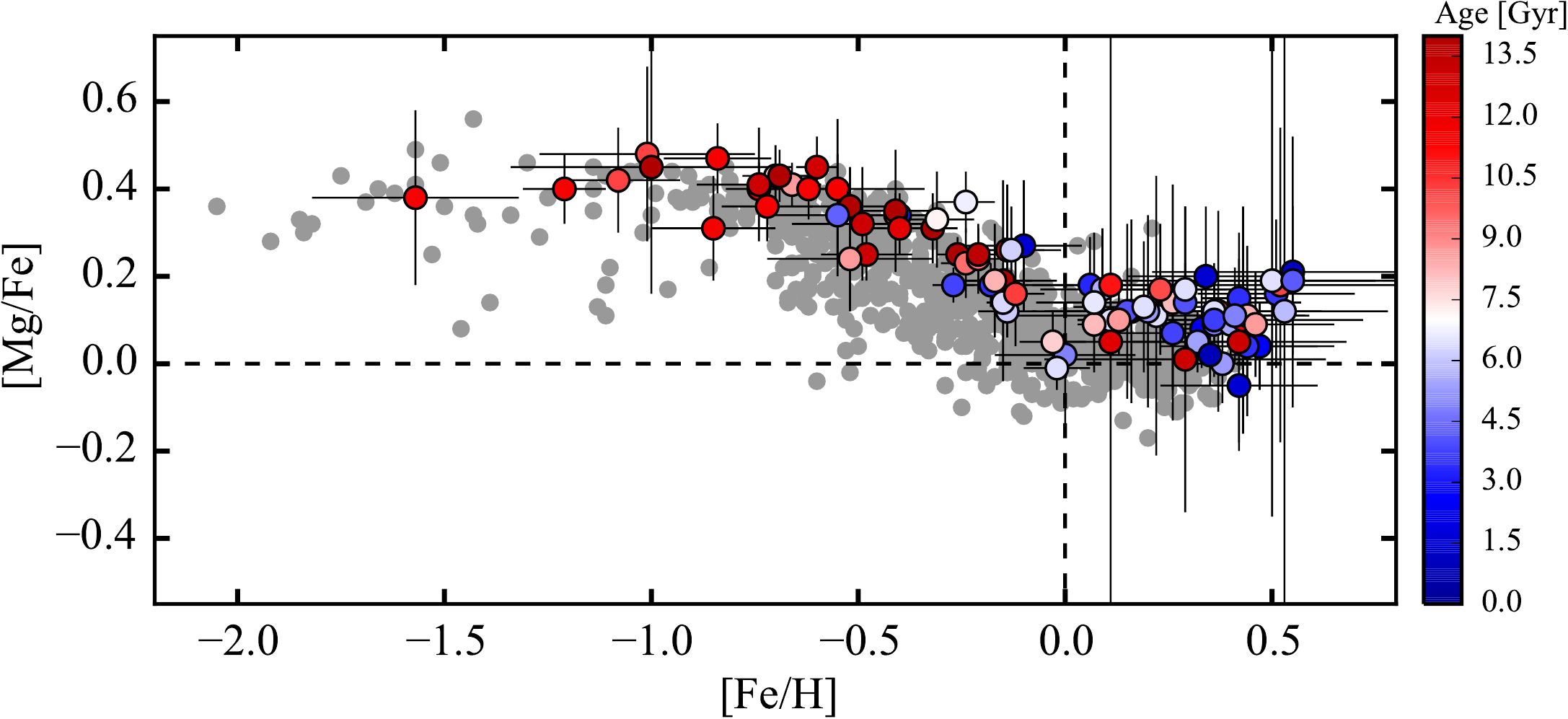}
\caption{From Figure 21 of \citet{2017arXiv170202971B}. The [Mg/Fe] vs [Fe/H] abundance trend for bulge stars (coloured points) are shown superimposed on the disk trends (grey points). The magnesium abundances are elevated with respect to the disk abundances at all [Fe/H] values. }
\label{Bensby2017}
\end{figure}

The relative abundance offsets between the bulge and the disk have not converged to zero as more data have come in. \citet{2017arXiv170202971B} compiled what is among the best datasets of bulge abundances, as they have high-resolution, high signal-to-noise abundances for 90 bulge stars located on the main-sequence turnoff and subgiant branch, analyzed using the same methods as their comparison disk sample, which is composed of stars in the solar neighbourhood. Some elements trace the same abundance trends as the disk, but magnesium, titanium, and aluminum do not. They typically trace the upper end of the larger distributions spanned by the thick and thin disks. The abundance trend of [Mg/Fe] vs [Fe/H] is shown in Figure \ref{Bensby2017}.

Relative abundance offsets between the bulge and the disk remain, which is a challenge to models of the Galaxy where the bulge is simply due to buckling thin and thick disks. As the abundance ratios are higher in magnesium in particular, suggesting a more rapid star formation, the suggestion is that the bulge formed faster than the disk, characteristic of an early, dissipative collapse. 


   \section{THE CLUMP-ORIGIN BULGE SCENARIO}
Before proceeding, a brief description of star-forming clumps will be given, though the description is itself a matter of active research. Whereas ``smooth exponential disks" are decent approximations to local disk galaxies and exact descriptions of many N-body models, star-forming galaxies at high-redshift are generally clumpy and gas-rich \citep{1995AJ....110.1576C,1996AJ....112..359V,2006Natur.442..786G}, with off-centre clumps accounting for 7\% of the stellar mass and 20\% of the star formation in massive, star-forming galaxies \citep{2012ApJ...753..114W}.

The clumps are regions of excess star-formation seen in the disks and proto-disks of high-redshift galaxies. \citet{2012MNRAS.422.3339W} studied the properties of eight clumps in three redshift $z \sim 1.3$ observed as part of the WiggleZ Dark Energy Survey. They had an average size of 1.5 Kpc and average Jeans mass of $4.2 \times 10^9 M_{\odot}$, and accounted for roughly half the stellar mass of the disks. 

Within this scenario, motivated by predictions from simulations \citep{1998Natur.392..253N,1999ApJ...514...77N}, some of these clumps will migrate to the centre of their galaxies due to dynamical friction, and thus form a bulge. 

The evidence presented here is that:
\begin{itemize}
\item High-redshift galaxies largely appear as clumpy galaxies, and thus this is the plausible set of ``initial" conditions for the Milky Way. 
\item The simulations succeed at predicting many of the observations.
\end{itemize}
This is less evidence than for the other two scenarios, but that is plausibly simply due to this being less researched topic. It is hoped that there will be further research testing whether or not the Milky Way bulge may be a clump-origin bulge. \citet{2014A&A...562A..66Z} also discussed the issue. They pointed out that the mean age of bulge stars corresponds to an epoch of gas-rich disks, with gas fractions sometimes exceeding 50\% \citep{2010Natur.463..781T,2010ApJ...713..686D}, which are more consistent with simulations of clumpy-galaxies than the usually gas-free N-body simulations of bar formation in disks. 

Readers interested in a more thorough review of bulge growth in high-redshift galaxies are referred to the excellent review by \citet{2016ASSL..418..355B}. In particular, Section 3, ``Mechanisms of bulge growth through high-redshift disk instabilities". 

\subsection{This is what high-redshift galaxies actually look like}

   \begin{figure*}
\begin{center}
\includegraphics[totalheight=0.27\textheight]{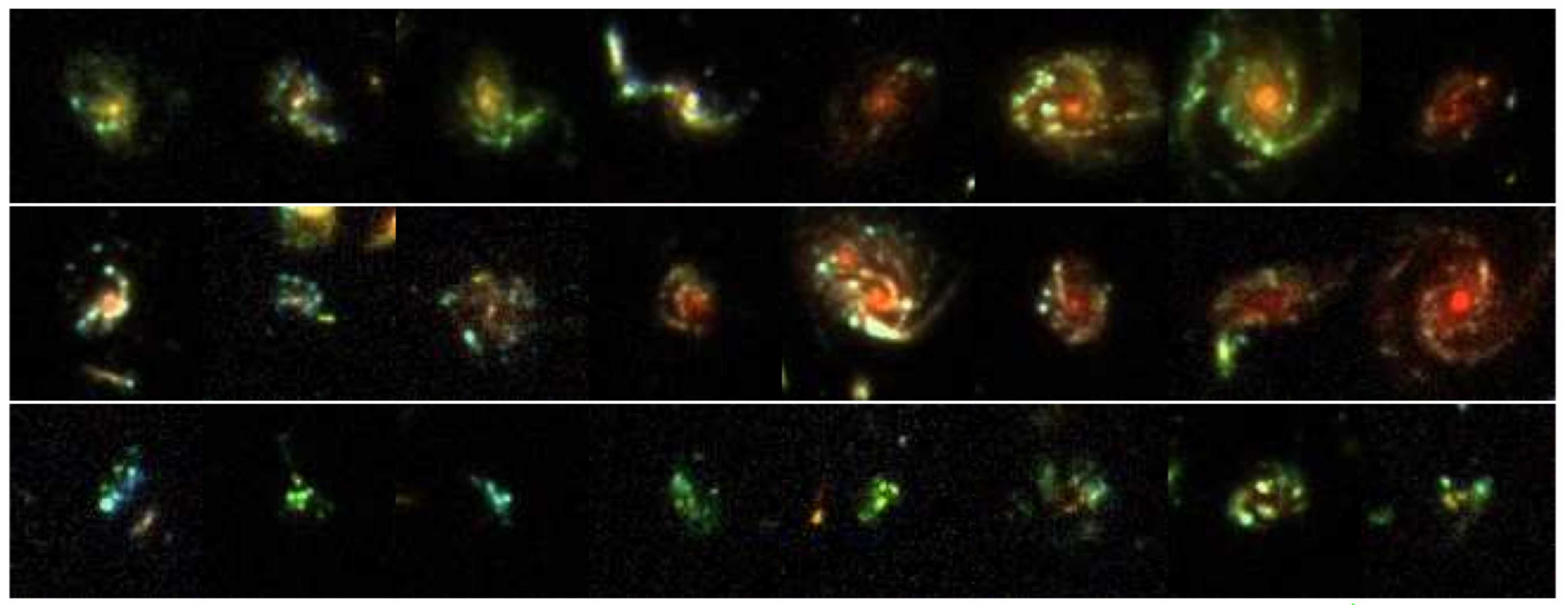}
\caption{Figure 4 from \citet{2015ApJ...800...39G} showing HST images of high-redshift galaxies. The norm is that of massive, star-forming clumps within disks, rather than simple and smooth exponential disks. }
\label{Guo2015}
\end{center}
\end{figure*}

\citet{2015ApJ...800...39G} analyzed 3,239 high-redshift, star-forming galaxies studied as part of the \textit{Cosmic Assembly Near-Infrared Deep Extragalactic Legacy Survey (CANDELS)}. They use a conservative definition for clumps -- a clump has to contribute at least 8\% of the UV light of a galaxy, which excludes smaller clumps. They also require clumps to be off-centre, which excludes central clumps and is thus limiting in our context as a central clump could obviously contribute to bulge formation. One of their mosaics of clumpy galaxies is shown in Figure \ref{Guo2015}.

They find that galaxies with $\log(M_{\star}./M_{\odot}) > 9.8$ have a 55\% probability of being clumpy at redshift $z \sim3$, down to 15\% at $z \sim0.5$, which is largely due to the fact star formation declines with decreasing redshift. At all redshifts, the clump contribution to rest-frame UV light peaks at $\log(M_{\star}./M_{\odot}) > 10.5$ -- the current stellar mass of the Milky Way.  Integrating over both clumpy and non-clumpy galaxies, they find that 4\%-10\% of the star formation takes place within these massive clumps. 

In other words, the clumpy galaxy is a very plausible assumption for the initial conditions of the Milky Way. Much has been said in this review that the pure disk galaxy can work, as \citet{2010ApJ...723...54K} has pointed out. However, the smooth and massive exponential disks are widely seen in observations of the local universe. That clumpy galaxies are the norm for high-redshift observations suggests that most massive galaxies have passed through a clumpy phase.

\subsection{Promising insights from simulations}
The prevalence of clumps in high-redshift galaxies renders them a legitimate point of discussion for the origin of the Milky Way.  

A way to test this is with comparison to the Milky Way, which \citet{2012MNRAS.422.1902I} did. They used an N-body/SPH model to study the evolution of an isolated disk galaxy where the clumps migrate to the centre via dynamical friction and form a clump-origin bulge. The final bulge resembles what they call a pseudo bulge, and is referred to as a bar throughout this paper and most of the Milky Way literature. The surface density profile is nearly exponential, the final shape of the bar is boxy, and the rotation is significant. The resulting stars are old and metal-rich, with a flat star-formation history in an interval of $\Delta t \sim 2$ Gyr followed by a rapid decline in star formation. They obtain a metallicity gradient that stretches across the full vertical extent of the bulge. All of these properties are qualitatively consistent with what is observed for the Milky Way bulge. However, these properties can be matched by other models, and further this is a comparison of a single clump-origin bulge model to those of the bulge. It is worthy of consideration, but it is far too premature to declare victory. 

One concern is that of whether or not the clumps actually do migrate to the centre, as they do the simulations previously discussed in this review. This is largely a question for theory, as the migration duration is too long for the baseline of observations. In the simulations of \citet{2012MNRAS.427..968H}, the inclusion of their prescription for stellar feedback disrupted the clumps, and prevented them from migrating to the bulge where they can coalesce. This prediction is not reproduced by the simulations of \citet{2014ApJ...780...57B}, who find that the ejection of stars in clumps due to stellar feedback is compensated by their accretion of gas from the gas-rich disks in which they are contained. 

\citet{2014MNRAS.443.3675M} studied 770 snapshots of 20 simulated galaxies using adaptive-mesh refinement cosmological models of Galaxy evolution. The global number of clumps were consistent with those in observations, an important check of the violent-disk instability hypothesis of clump formation. They did not study the properties of the final resulting bulge, but did say that if the clumps can survive accretion onto the centre of the galaxy, they are expected to accrete gas from the surrounding interstellar medium (similarly to \citealt{2014ApJ...780...57B}), and will thus show gradients in their mean properties  with respect to separation from the centre of their galaxy, such as those measured by \citet{2011ApJ...739...45F} that clumps closer to the centres of their disks are redder, older, and more massive. Further, they found in their simulations was that a full 91\% of galaxies develop a bulge clump. These are massive, typically equivalent to 40\% of the disk mass, with 20\% of the star formation, and gas fractions of less than 1\%. 

\citet{2017MNRAS.464..635M} study 34 galaxies with more sophisticated prescriptions. Among their findings, they find that the inclusion of radiation pressure disrupts the smaller clumps, reducing their lifetimes to a few free-fall times, but that the more massive and dense clumps still nevertheless survive and migrate to the centre. The inclusion of radiation pressure reduces the number of long-lived clumps by 81\%.  Radiation pressure has little to no effect on the bulge clumps, with $\sim$83\% of simulated galaxies hosting a bulge clump.

\section{A CLUE AS TO THE ORIGINS OF THE BULGE FROM APOGEE}
A significant clue as to the origin of the metal-poor stars in the bulge has been identified by the APOGEE collaboration. Previously, the detailed chemical abundance trends for the bulge have only been interpreted in their mean, due to the large observational error. The mean [X/Fe] vs [Fe/H] can and has be compared to the thin disk, thick disk, and halo, but the scatter has not yet been of particular insight. 

\citet{2017MNRAS.465..501S}  found a population of nitrogen-rich stars in the bulge, predominantly at [Fe/H] $\leq -1.0$. These stars have enhanced nitrogen, aluminum, and depleted carbon, characteristic of the ``second-generation" stars in globular clusters \citep{2009A&A...505..117C}. Given that even surviving globular clusters must have been far more massive at birth \citep{2012ApJ...758...21C} to produce their second generation, this suggests that between 50\% and 100\% of bulge stars with [Fe/H] $\leq -1.0$ formed in disassociated globular clusters. 

This is a clue to the origin of the bulge, but it is not clear which line of evidence it can be used to support. That is why it is left as a separate section. 

\section{DISCUSSION AND CONCLUSION}
\label{sec:Conclusion}
The level of inputs new to the last decade, both observational and theoretical, that can inform and constrain bulge formation scenarios is truly spectacular.  Global photometric maps are now available from the optical through to the mid-infrared, with substantial coverage in the variability domaine. Spectroscopic data sets are now available toward a large fraction of bulge stars, and toward the full metallicity range. Knowledge of what high-redshift galaxies look like, including plausible Milky Way precursors, is greater than it's ever been. The breadth and depth of models if constantly increasing. 

It would be tempting to say that the situation remains one of uncertainty between different scenarios, but that would be so limiting as to be inaccurate. In a competition between the buckling disk and the classical bulge, the buckling disk is winning. The peanut/X-shape, the long bar, the correlation between the mean and skewness of the velocity distribution functions, and so on are non-trivial predictions that are required of any theory of bulge formation and at this time require a buckling disk. The classical bulge may dominate for the 5\% of stars with [Fe/H] $\lesssim -1.0$ and a minority of more metal-rich stars, but that is an upper limit on its contribution. 

The clump-origin bulge scenario may prove to be a viable alternative. It cannot be ignored given the ubiquity of star-forming, gas-rich clumps in high-redshift galaxies. It is more likely than not that the Milky Way was, at one time, a clumpy galaxy. More simulations and comparisons of said simulations to observations  are needed to ascertain whether or not this describes the bulge assembly history. One plausible hybrid scenario, is if the clumpy phase of the Milky Way led to the thick disk \citep{2014MNRAS.441..243I}, with the thick disk predominantly responsible for bulge stars with $-1.0 \lesssim \rm{[Fe/H]} \lesssim -0.50$ \citep{2015A&A...577A...1D}. 

One thing is certain, the Milky Way bulge is a sensitive probe of Galactic assembly history, and research of its properties will continue yielding insights thereof.

\section*{Acknowledgments}
DMN was supported by the Allan C. and Dorothy H. Davis Fellowship. Chiaki Kobayashi, Rosemary Wyse, and Alex de la Vega are thanked for helpful discussions.


\begin{thebibliography}{}
\bibitem[Alves-Brito et al.(2010)]{2010A&A...513A..35A} Alves-Brito, A., Mel{\'e}ndez, J., Asplund, M., Ram{\'{\i}}rez, I., \& Yong, D.\ 2010, \aap, 513, A35 
\bibitem[Athanassoula \& Misiriotis(2002)]{2002MNRAS.330...35A} Athanassoula, E., \& Misiriotis, A.\ 2002, \mnras, 330, 35 
\bibitem[Athanassoula(2003)]{2003MNRAS.341.1179A} Athanassoula, E.\ 2003, \mnras, 341, 1179 
\bibitem[Athanassoula(2005)]{2005MNRAS.358.1477A} Athanassoula, E.\ 2005, \mnras, 358, 1477 
\bibitem[Athanassoula(2012)]{2012EPJWC..1906004A} Athanassoula, E.\ 2012, European Physical Journal Web of Conferences, 19, 06004 
\bibitem[Athanassoula et al.(2013)]{2013MNRAS.429.1949A} Athanassoula, E., Machado, R.~E.~G., \& Rodionov, S.~A.\ 2013, \mnras, 429, 1949 
\bibitem[Athanassoula et al.(2017)]{2017MNRAS.467L..46A} Athanassoula, E., Rodionov, S.~A., \& Prantzos, N.\ 2017, \mnras, 467, L46 
\bibitem[Benjamin et al.(2005)]{2005ApJ...630L.149B} Benjamin, R.~A., Churchwell, E., Babler, B.~L., et al.\ 2005, \apjl, 630, L149 
\bibitem[Barbuy et al.(2006)]{2006A&A...449..349B} Barbuy, B., Zoccali, M., Ortolani, S., et al.\ 2006, \aap, 449, 349 
\bibitem[Barbuy et al.(2007)]{2007AJ....134.1613B} Barbuy, B., Zoccali, M., Ortolani, S., et al.\ 2007, \aj, 134, 1613 
\bibitem[Barbuy et al.(2009)]{2009A&A...507..405B} Barbuy, B., Zoccali, M., Ortolani, S., et al.\ 2009, \aap, 507, 405 
\bibitem[Bensby et al.(2017)]{2017arXiv170202971B} Bensby, T., Feltzing, S., Gould, A., et al.\ 2017, arXiv:1702.02971 
\bibitem[Bournaud et al.(2014)]{2014ApJ...780...57B} Bournaud, F., Perret, V., Renaud, F., et al.\ 2014, \apj, 780, 57 
\bibitem[Bournaud(2016)]{2016ASSL..418..355B} Bournaud, F.\ 2016, Galactic Bulges, 418, 355 
\bibitem[Brown et al.(2010)]{2010ApJ...725L..19B} Brown, T.~M., Sahu, K., Anderson, J., et al.\ 2010, \apjl, 725, L19 
\bibitem[Cabrera-Lavers et al.(2007)]{2007A&A...465..825C} Cabrera-Lavers, A., Hammersley, P.~L., Gonz{\'a}lez-Fern{\'a}ndez, C., et al.\ 2007, \aap, 465, 825 
\bibitem[Cabrera-Lavers et al.(2008)]{2008A&A...491..781C} Cabrera-Lavers, A., Gonz{\'a}lez-Fern{\'a}ndez, C., Garz{\'o}n, F., Hammersley, P.~L., \& L{\'o}pez-Corredoira, M.\ 2008, \aap, 491, 781 
\bibitem[Cao et al.(2013)]{2013MNRAS.434..595C} Cao, L., Mao, S., Nataf, D., Rattenbury, N.~J., \& Gould, A.\ 2013, \mnras, 434, 595 
\bibitem[Carlberg(1984)]{1984ApJ...286..403C} Carlberg, R.~G.\ 1984, \apj, 286, 403 
\bibitem[Carretta et al.(2009)]{2009A&A...505..117C} Carretta, E., Bragaglia, A., Gratton, R.~G., et al.\ 2009, \aap, 505, 117 
\bibitem[Clarkson et al.(2011)]{2011ApJ...735...37C} Clarkson, W.~I., Sahu, K.~C., Anderson, J., et al.\ 2011, \apj, 735, 37 
\bibitem[Combes \& Sanders(1981)]{1981A&A....96..164C} Combes, F., \& Sanders, R.~H.\ 1981, \aap, 96, 164 
\bibitem[Combes(2000)]{2000bgfp.conf..413C} Combes, F.\ 2000, Building Galaxies; from the Primordial Universe to the Present, 413 
\bibitem[Conroy(2012)]{2012ApJ...758...21C} Conroy, C.\ 2012, \apj, 758, 21 
\bibitem[Cowie et al.(1995)]{1995AJ....110.1576C} Cowie, L.~L., Hu, E.~M., \& Songaila, A.\ 1995, \aj, 110, 1576 
\bibitem[Daddi et al.(2010)]{2010ApJ...713..686D} Daddi, E., Bournaud, F., Walter, F., et al.\ 2010, \apj, 713, 686 
\bibitem[Debattista et al.(2016)]{2016arXiv161109023D} Debattista, V.~P., Ness, M., Gonzalez, O.~A., et al.\ 2016, arXiv:1611.09023 
\bibitem[Delgado Mena et al.(2017)]{2017arXiv170504349D} Delgado Mena, E., Tsantaki, M., Adibekyan, V.~Z., et al.\ 2017, arXiv:1705.04349 
\bibitem[D{\'e}k{\'a}ny et al.(2013)]{2013ApJ...776L..19D} D{\'e}k{\'a}ny, 
I., Minniti, D., Catelan, M., et al.\ 2013, \apjl, 776, L19 
\bibitem[Di Matteo et al.(2015)]{2015A&A...577A...1D} Di Matteo, P., G{\'o}mez, A., Haywood, M., et al.\ 2015, \aap, 577, A1 
\bibitem[Di Matteo(2016)]{2016PASA...33...27D} Di Matteo, P.\ 2016, \pasa, 33, e027 
\bibitem[Dotter et al.(2010)]{2010ApJ...708..698D} Dotter, A., Sarajedini, A., Anderson, J., et al.\ 2010, \apj, 708, 698 
\bibitem[Dwek et al.(1995)]{1995ApJ...445..716D} Dwek, E., Arendt, R.~G., Hauser, M.~G., et al.\ 1995, \apj, 445, 716 
\bibitem[Dwek(2004)]{2004ApJ...611L.109D} Dwek, E.\ 2004, \apjl, 611, L109 
\bibitem[Eggen et al.(1962)]{1962ApJ...136..748E} Eggen, O.~J., Lynden-Bell, D., \& Sandage, A.~R.\ 1962, \apj, 136, 748 
\bibitem[F{\"o}rster Schreiber et al.(2011)]{2011ApJ...739...45F} F{\"o}rster Schreiber, N.~M., Shapley, A.~E., Genzel, R., et al.\ 2011, \apj, 739, 45 
\bibitem[Freeman et al.(2013)]{2013MNRAS.428.3660F} Freeman, K., Ness, M., Wylie-de-Boer, E., et al.\ 2013, \mnras, 428, 3660 
\bibitem[Fulbright et al.(2007)]{2007ApJ...661.1152F} Fulbright, J.~P., McWilliam, A., \& Rich, R.~M.\ 2007, \apj, 661, 1152 
\bibitem[Gardner et al.(2014)]{2014MNRAS.438.3275G} Gardner, E., Debattista, V.~P., Robin, A.~C., V{\'a}squez, S., \& Zoccali, M.\ 2014, \mnras, 438, 3275 
\bibitem[Genzel et al.(2006)]{2006Natur.442..786G} Genzel, R., Tacconi, L.~J., Eisenhauer, F., et al.\ 2006, \nat, 442, 786 
\bibitem[Gonzalez et al.(2013)]{2013A&A...552A.110G} Gonzalez, O.~A., Rejkuba, M., Zoccali, M., et al.\ 2013, \aap, 552, A110 
\bibitem[Guo et al.(2015)]{2015ApJ...800...39G} Guo, Y., Ferguson, H.~C., Bell, E.~F., et al.\ 2015, \apj, 800, 39 
\bibitem[Hohl \& Zang(1979)]{1979AJ.....84..585H} Hohl, F., \& Zang, T.~A.\ 1979, \aj, 84, 585 
\bibitem[Hopkins et al.(2012)]{2012MNRAS.427..968H} Hopkins, P.~F., Kere{\v s}, D., Murray, N., Quataert, E., \& Hernquist, L.\ 2012, \mnras, 427, 968 
\bibitem[Howes et al.(2014)]{2014MNRAS.445.4241H} Howes, L.~M., Asplund, M., Casey, A.~R., et al.\ 2014, \mnras, 445, 4241 
\bibitem[Inoue \& Saitoh(2012)]{2012MNRAS.422.1902I} Inoue, S., \& Saitoh, T.~R.\ 2012, \mnras, 422, 1902 
\bibitem[Inoue \& Saitoh(2014)]{2014MNRAS.441..243I} Inoue, S., \& Saitoh, T.~R.\ 2014, \mnras, 441, 243 
\bibitem[Johnson et al.(2011)]{2011ApJ...732..108J} Johnson, C.~I., Rich, R.~M., Fulbright, J.~P., Valenti, E., \& McWilliam, A.\ 2011, \apj, 732, 108 
\bibitem[Kauffmann et al.(1993)]{1993MNRAS.264..201K} Kauffmann, G., White, S.~D.~M., \& Guiderdoni, B.\ 1993, \mnras, 264, 201 
\bibitem[Kobayashi \& Nakasato(2011)]{2011ApJ...729...16K} Kobayashi, C., \& Nakasato, N.\ 2011, \apj, 729, 16 
\bibitem[Kormendy et al.(2009)]{2009ApJS..182..216K} Kormendy, J., Fisher, D.~B., Cornell, M.~E., \& Bender, R.\ 2009, \apjs, 182, 216 
\bibitem[Kormendy et al.(2010)]{2010ApJ...723...54K} Kormendy, J., Drory, N., Bender, R., \& Cornell, M.~E.\ 2010, \apj, 723, 54
\bibitem[Kunder et al.(2012)]{2012AJ....143...57K} Kunder, A., Koch, A., Rich, R.~M., et al.\ 2012, \aj, 143, 57 
\bibitem[Kunder et al.(2015)]{2015ApJ...808L..12K} Kunder, A., Rich, R.~M., 
Hawkins, K., et al.\ 2015, \apjl, 808, L12 
\bibitem[Kunder et al.(2016)]{2016ApJ...821L..25K} Kunder, A., Rich, R.~M., Koch, A., et al.\ 2016, \apjl, 821, L25 
\bibitem[Lee(1992)]{1992AJ....104.1780L} Lee, Y.-W.\ 1992, \aj, 104, 1780 
\bibitem[Li \& Shen(2012)]{2012ApJ...757L...7L} Li, Z.-Y., \& Shen, J.\ 2012, \apjl, 757, L7 
\bibitem[Lucas et al.(2008)]{2008MNRAS.391..136L} Lucas, P.~W., Hoare, M.~G., Longmore, A., et al.\ 2008, \mnras, 391, 136 
\bibitem[Mandelker et al.(2014)]{2014MNRAS.443.3675M} Mandelker, N., Dekel, A., Ceverino, D., et al.\ 2014, \mnras, 443, 3675 
\bibitem[Mandelker et al.(2017)]{2017MNRAS.464..635M} Mandelker, N., Dekel, A., Ceverino, D., et al.\ 2017, \mnras, 464, 635 
\bibitem[Martinez-Valpuesta et al.(2006)]{2006ApJ...637..214M} Martinez-Valpuesta, I., Shlosman, I., \& Heller, C.\ 2006, \apj, 637, 214 
\bibitem[Martinez-Valpuesta \& Gerhard(2011)]{2011ApJ...734L..20M} Martinez-Valpuesta, I., \& Gerhard, O.\ 2011, \apjl, 734, L20 
\bibitem[Martinez-Valpuesta \& Gerhard(2013)]{2013ApJ...766L...3M} Martinez-Valpuesta, I., \& Gerhard, O.\ 2013, \apjl, 766, L3 
\bibitem[Matteucci \& Greggio(1986)]{1986A&A...154..279M} Matteucci, F., \& Greggio, L.\ 1986, \aap, 154, 279 
\bibitem[McWilliam \& Rich(1994)]{1994ApJS...91..749M} McWilliam, A., \& Rich, R.~M.\ 1994, \apjs, 91, 749 
\bibitem[McWilliam 
\& Zoccali(2010)]{2010ApJ...724.1491M} McWilliam, A., \& Zoccali, M.\ 2010, \apj, 724, 1491 
\bibitem[Melvin et al.(2014)]{2014MNRAS.438.2882M} Melvin, T., Masters, K., Lintott, C., et al.\ 2014, \mnras, 438, 2882 
\bibitem[Miller(1978)]{1978ApJ...223..811M} Miller, R.~H.\ 1978, \apj, 223, 811 
\bibitem[Miller \& Smith(1979)]{1979ApJ...227..785M} Miller, R.~H., \& Smith, B.~F.\ 1979, \apj, 227, 785 
\bibitem[Minniti et al.(1995)]{1995MNRAS.277.1293M} Minniti, D., Olszewski, E.~W., Liebert, J., et al.\ 1995, \mnras, 277, 1293 
\bibitem[Nataf et al.(2010)]{2010ApJ...721L..28N} Nataf, D.~M., Udalski, 
A., Gould, A., Fouqu{\'e}, P., \& Stanek, K.~Z.\ 2010, \apjl, 721, L28 
\bibitem[Nataf et al.(2015)]{2015MNRAS.447.1535N} Nataf, D.~M., Udalski, A., Skowron, J., et al.\ 2015, \mnras, 447, 1535 
\bibitem[Nataf et al.(2013)]{2013ApJ...769...88N} Nataf, D.~M., Gould, A., Fouqu{\'e}, P., et al.\ 2013, \apj, 769, 88 
\bibitem[Ness et al.(2012)]{2012ApJ...756...22N} Ness, M., Freeman, K., Athanassoula, E., et al.\ 2012, \apj, 756, 22 
\bibitem[Ness et al.(2013)]{2013MNRAS.430..836N} Ness, M., Freeman, K., Athanassoula, E., et al.\ 2013, \mnras, 430, 836 
\bibitem[Ness et al.(2013)]{2013MNRAS.432.2092N} Ness, M., Freeman, K., Athanassoula, E., et al.\ 2013, \mnras, 432, 2092 
\bibitem[Ness \& Lang(2016)]{2016AJ....152...14N} Ness, M., \& Lang, D.\ 2016, \aj, 152, 14 
\bibitem[Ness et al.(2016)]{2016ApJ...819....2N} Ness, M., Zasowski, G., Johnson, J.~A., et al.\ 2016, \apj, 819, 2 
\bibitem[Noguchi(1998)]{1998Natur.392..253N} Noguchi, M.\ 1998, \nat, 392, 253 
\bibitem[Noguchi(1999)]{1999ApJ...514...77N} Noguchi, M.\ 1999, \apj, 514, 77 
\bibitem[Paczy{\'n}ski 
\& Stanek(1998)]{1998ApJ...494L.219P} Paczy{\'n}ski, B., \& Stanek, K.~Z.\ 1998, \apjl, 494, L219 
\bibitem[P{\'e}rez-Villegas et al.(2017)]{2017MNRAS.464L..80P} P{\'e}rez-Villegas, A., Portail, M., \& Gerhard, O.\ 2017, \mnras, 464, L80 
\bibitem[Pietrukowicz et al.(2012)]{2012ApJ...750..169P} Pietrukowicz, P., 
Udalski, A., Soszy{\'n}ski, I., et al.\ 2012, \apj, 750, 169 
\bibitem[Pietrukowicz et al.(2015)]{2015ApJ...811..113P} Pietrukowicz, P., 
Koz{\l}owski, S., Skowron, J., et al.\ 2015, \apj, 811, 113 
\bibitem[Planck Collaboration et al.(2014)]{2014A&A...571A..16P} Planck Collaboration, Ade, P.~A.~R., Aghanim, N., et al.\ 2014, \aap, 571, A16 
\bibitem[Portail et al.(2015a)]{2015MNRAS.450L..66P} Portail, M., Wegg, C., \& Gerhard, O.\ 2015, \mnras, 450, L66 
\bibitem[Portail et al.(2015b)]{2015MNRAS.448..713P} Portail, M., Wegg, C., Gerhard, O., \& Martinez-Valpuesta, I.\ 2015, \mnras, 448, 713 
\bibitem[Portail et al.(2017)]{2017MNRAS.465.1621P} Portail, M., Gerhard, O., Wegg, C., \& Ness, M.\ 2017, \mnras, 465, 1621 
\bibitem[Quillen et al.(2014)]{2014MNRAS.437.1284Q} Quillen, A.~C., Minchev, I., Sharma, S., Qin, Y.-J., \& Di Matteo, P.\ 2014, \mnras, 437, 1284 
\bibitem[Rattenbury et al.(2007)]{2007MNRAS.378.1064R} Rattenbury, N.~J., Mao, S., Sumi, T., \& Smith, M.~C.\ 2007, \mnras, 378, 1064 
\bibitem[Rojas-Arriagada et 
al.(2014)]{2014A&A...569A.103R} Rojas-Arriagada, A., Recio-Blanco, A., Hill, V., et al.\ 2014, \aap, 569, A103 
\bibitem[Raha et al.(1991)]{1991Natur.352..411R} Raha, N., Sellwood, J.~A., James, R.~A., \& Kahn, F.~D.\ 1991, \nat, 352, 411 
\bibitem[Rich et al.(2007)]{2007ApJ...658L..29R} Rich, R.~M., Reitzel, D.~B., Howard, C.~D., \& Zhao, H.\ 2007, \apjl, 658, L29 
\bibitem[Rojas-Arriagada et al.(2014)]{2014A&A...569A.103R} Rojas-Arriagada, A., Recio-Blanco, A., Hill, V., et al.\ 2014, \aap, 569, A103 
\bibitem[Saha et al.(2012)]{2012MNRAS.421..333S} Saha, K., Martinez-Valpuesta, I., \& Gerhard, O.\ 2012, \mnras, 421, 333 
\bibitem[Saha \& Gerhard(2013)]{2013MNRAS.430.2039S} Saha, K., \& Gerhard, O.\ 2013, \mnras, 430, 2039 
\bibitem[Saha et al.(2016)]{2016A&A...588A..42S} Saha, K., Gerhard, O., \& Martinez-Valpuesta, I.\ 2016, \aap, 588, A42 
\bibitem[Saito et al.(2012)]{2012A&A...537A.107S} Saito, R.~K., Hempel, M., Minniti, D., et al.\ 2012, \aap, 537, A107 
\bibitem[Schiavon et al.(2017)]{2017MNRAS.465..501S} Schiavon, R.~P., Zamora, O., Carrera, R., et al.\ 2017, \mnras, 465, 501 
\bibitem[Shen et al.(2010)]{2010ApJ...720L..72S} Shen, J., Rich, R.~M., Kormendy, J., et al.\ 2010, \apjl, 720, L72 
\bibitem[S{\'a}nchez et al.(2014)]{2014A&A...563A..49S} S{\'a}nchez, S.~F., Rosales-Ortega, F.~F., Iglesias-P{\'a}ramo, J., et al.\ 2014, \aap, 563, A49 
\bibitem[Simmons et al.(2014)]{2014MNRAS.445.3466S} Simmons, B.~D., Melvin, T., Lintott, C., et al.\ 2014, \mnras, 445, 3466 
\bibitem[Skokos et al.(2002)]{2002MNRAS.333..847S} Skokos, C., Patsis, P.~A., \& Athanassoula, E.\ 2002, \mnras, 333, 847 
\bibitem[Skrutskie et al.(2006)]{2006AJ....131.1163S} Skrutskie, M.~F., Cutri, R.~M., Stiening, R., et al.\ 2006, \aj, 131, 1163 
\bibitem[Stanek et al.(1997)]{1997ApJ...477..163S} Stanek, K.~Z., Udalski, A., Szyma{\'N}ski, M., et al.\ 1997, \apj, 477, 163 
\bibitem[Syer \& Tremaine(1996)]{1996MNRAS.282..223S} Syer, D., \& Tremaine, S.\ 1996, \mnras, 282, 223 
\bibitem[Tacconi et al.(2010)]{2010Natur.463..781T} Tacconi, L.~J., Genzel, R., Neri, R., et al.\ 2010, \nat, 463, 781 
\bibitem[Tiede et al.(1995)]{1995AJ....110.2788T} Tiede, G.~P., Frogel, 
J.~A., \& Terndrup, D.~M.\ 1995, \aj, 110, 2788 
\bibitem[Tumlinson(2010)]{2010ApJ...708.1398T} Tumlinson, J.\ 2010, \apj, 708, 1398 
\bibitem[Udalski et al.(2008)]{2008AcA....58...69U} Udalski, A., Szymanski, 
M.~K., Soszynski, I., \& Poleski, R.\ 2008, \actaa, 58, 69 
\bibitem[Udalski et al.(2015)]{2015AcA....65....1U} Udalski, A., Szyma{\'n}ski, M.~K., \& Szyma{\'n}ski, G.\ 2015, \actaa, 65, 1 
\bibitem[van den Bergh et al.(1996)]{1996AJ....112..359V} van den Bergh, S., Abraham, R.~G., Ellis, R.~S., et al.\ 1996, \aj, 112, 359 
\bibitem[V{\'a}squez et al.(2013)]{2013A&A...555A..91V} V{\'a}squez, S., Zoccali, M., Hill, V., et al.\ 2013, \aap, 555, A91 
\bibitem[Wegg 
\& Gerhard(2013)]{2013MNRAS.435.1874W} Wegg, C., \& Gerhard, O.\ 2013, \mnras, 435, 1874 
\bibitem[Wegg et al.(2015)]{2015MNRAS.450.4050W} Wegg, C., Gerhard, O., \& Portail, M.\ 2015, \mnras, 450, 4050 
\bibitem[White \& Rees(1978)]{1978MNRAS.183..341W} White, S.~D.~M., \& Rees, M.~J.\ 1978, \mnras, 183, 341 
\bibitem[White(1980)]{1980MNRAS.191P...1W} White, S.~D.~M.\ 1980, \mnras, 191, 1P 
\bibitem[Wisnioski et al.(2012)]{2012MNRAS.422.3339W} Wisnioski, E., Glazebrook, K., Blake, C., et al.\ 2012, \mnras, 422, 3339 
\bibitem[Wright(2006)]{2006PASP..118.1711W} Wright, E.~L.\ 2006, \pasp, 118, 1711 
\bibitem[Wright et al.(2010)]{2010AJ....140.1868W} Wright, E.~L., Eisenhardt, P.~R.~M., Mainzer, A.~K., et al.\ 2010, \aj, 140, 1868-1881 
\bibitem[Wuyts et al.(2012)]{2012ApJ...753..114W} Wuyts, S., F{\"o}rster Schreiber, N.~M., Genzel, R., et al.\ 2012, \apj, 753, 114 
\bibitem[Zasowski et al.(2016)]{2016ApJ...832..132Z} Zasowski, G., Ness, M.~K., Garc{\'{\i}}a P{\'e}rez, A.~E., et al.\ 2016, \apj, 832, 132 
\bibitem[Zoccali et al.(2008)]{2008A&A...486..177Z} Zoccali, M., Hill, V., Lecureur, A., et al.\ 2008, \aap, 486, 177 
\bibitem[Zoccali et al.(2014)]{2014A&A...562A..66Z} Zoccali, M., Gonzalez, O.~A., Vasquez, S., et al.\ 2014, \aap, 562, A66 
\bibitem[Zoccali et al.(2017)]{2017A&A...599A..12Z} Zoccali, M., Vasquez, S., Gonzalez, O.~A., et al.\ 2017, \aap, 599, A12 
\end{thebibliography}
\end{document}